\documentclass[journal=jpclcd,manuscript=article]{achemso}

\usepackage[version=3]{mhchem} 
\usepackage{mciteplus}
\usepackage{xcolor}

\author{Harish Gudla}
\author{Yunqi Shao}
\author{Supho Phunnarungsi}
\author{Daniel Brandell}
\author{Chao Zhang}
\email{chao.zhang@kemi.uu.se}
\affiliation[UU]
{Department of Chemistry-\AA{}ngstr\"{o}m Laboratory, Uppsala University, L\"{a}gerhyddsv\"{a}gen 1, BOX 538, 75121, Uppsala, Sweden}

\title[Dissect-Ion]
  {Importance of the Ion-Pair Lifetime in Polymer Electrolytes}

\begin{document}

\begin{tocentry}
  \includegraphics[width=1\linewidth]{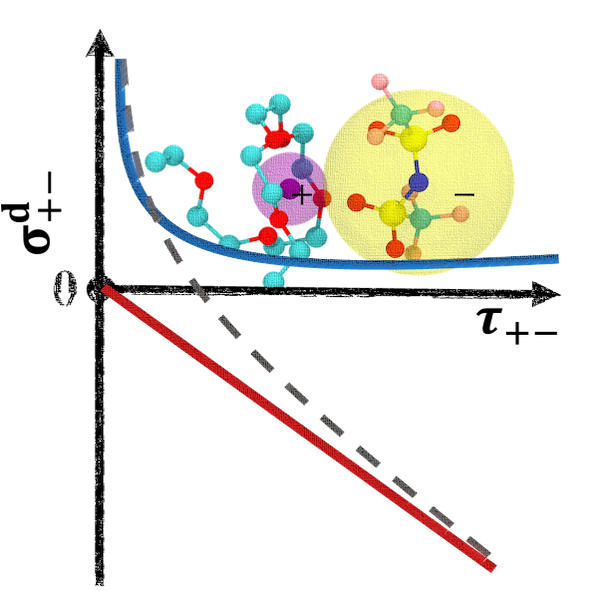}
  \label{For Table of Contents Only}
\end{tocentry}

\begin{abstract}
  Ion-pairing is commonly considered as a culprit for the reduced ionic conductivity in polymer electrolyte systems. However, this simple thermodynamic picture should not be taken literally, as ion-pairing is a dynamical phenomenon.  Here we construct model PEO–LiTFSI systems with different degree of ion-pairing by tuning solvent polarity, and examine the relation between the cation-anion distinct conductivity $\sigma^\textrm{d}_{+-}$ and the lifetime of ion-pairs $\tau_{+-}$ using molecular dynamics simulations. It is found that there exist two distinct regimes where  $\sigma^\textrm{d}_{+-}$ scales with $1/\tau_{+-}$ and $\tau_{+-}$ respectively, and the latter is a signature of longer-lived ion-pairs which contribute negatively to the total ionic conductivity.  This suggests that ion-pairs are kinetically different depending on the solvent polarity, which renders the ion-pair lifetime highly important when discussing its effect on ion transport in polymer electrolyte systems. 
\end{abstract}

Ion-pairing in electrolyte solutions~\cite{Marcus:2006bw, Macchioni:2005bu, Rehm:2010kp, Brak:2012dc, NavNidhiRajput:2015jz,Huang:2020de} results from a delicate balance between ion-solvent and ion-ion interactions. One common approach to define an ion-pair is using Bjerrum's criterion~\cite{Bjerrum:1926vx}, in which the distance $r_{+-}$ is smaller than the effective range $-q_+q_-/2\varepsilon k_bT$ (half of the Bjerrum length) with $\varepsilon$ as the dielectric constant of the solution,  $q_+$ and $q_-$ being the ionic charges, the Boltzmann constant $k_b$, and the temperature $T$. Bjerrum's criterion suggests that the solvent polarity plays a critical role in the formation of ion-pairs, rendering a distinction between contact ion-pairs (CIPs) and solvent separated ion-pairs (SSIPs)~\cite{Marcus:2006bw}. In addition, it implies that the formation of pairs of equal ionic species is unlikely to occur due to the electrostatic repulsion, but that the possibility of forming triplets or larger aggregates, e.g. a anion-cation-anion cluster, cannot be excluded~\cite{Evans:1987ho,Villaluenga:2018br}.

The idea that ion-pairing affects the ionic conductivity was introduced early on by Arrhenius, who ascribed the decrease of the equivalent conductivity at a higher concentration to the formation of charge-neutral ion-pairs~\cite{Fawcett:2004ww}. This idea has been put forward using the molar conductivity ratio $\Lambda_\textrm{EIS}/\Lambda_\textrm{NMR}$ measured by electrochemical impedance spectroscopy (EIS) and pulse-field gradient NMR to quantify the ionicity (the degree of dissociativity), particularly for ionic liquids~\cite{MacFarlane:2009ba,Ueno:2010kg} and polymer electrolytes~\cite{Stolwijk:2013do}. Nevertheless, it has been realized that deviations of the ionic conductivity from the Nernst-Einstein relation cannot solely be attributed to the formation of ion-pairs~\cite{harris_relationsfractionalstokes_2010,KashyapAnnapureddyEtAl_2011,2015_KirchnerMalbergEtAl}, where other factors such as the hydrodynamic interactions manifested via viscosity can play an important role.~\cite{ShaoShigenobuEtAl_2020}

To describe the effect of ion-pairing on the ionic conductivity, one needs an observable which can be accessed both theoretically and experimentally. The key quantity used here is the cation-anion distinct conductivity $\sigma^\textrm{d}_{+-}$ from liquid state theory~\cite{Hansen:2014uv,Zhong:1988co}:
\begin{equation}
  \label{eq:distinct conductivity}
  \sigma^\textrm{d}_{+-} = \lim_{t \rightarrow \infty} \frac{1}{3tk_bT\Omega} \left[
  \sum_{i, +}  \sum_{j, -} \langle
  q_{i,+}q_{j,-}\Delta\mathbf{r}_{i,+}(t) \cdot \Delta\mathbf{r}_{j,-}(t)
  \rangle \right] \\
\end{equation}
where $\Omega$ is the volume of the system and $\Delta\mathbf{r}(t)$ is the displacement vector of each ion at time $t$. Note that $\sigma^\textrm{d}_{+-}$ is experimentally measurable~\cite{Woolf:1978db, VargasBarbosa:2020eg} and directly related to the Onsager transport coefficient $\Omega_{+-}$~\cite{Zhong:1988co,Fong:2021ft}.  

Somewhat unexpectedly,  $\sigma^\textrm{d}_{+-}$  is often found positive (instead of negative as in Arrhenius' picture) in different types of electrolyte systems, spanning categories from aqueous electrolyte solutions to polymer ionic liquids~\cite{KashyapAnnapureddyEtAl_2011,McDaniel:2018du,Li:2019jc,Shao:2020gp, ShaoShigenobuEtAl_2020, Fong:2020df, Pfeifer:2021ei}. This suggests that the existence of ion-pairs, as evinced by a number of spectroscopic experiments,~\cite{Park:2010fz, Stange:2013kc,Chaurasia:2013kp} does not necessarily imply a negative contribution to the measured ionic conductivity but can instead contribute to an increase in the transport of ions. Therefore, understanding the ion-pairing effect on polymer electrolytes is crucial, as their application in energy storage systems is largely limited by a low ionic conductivity.~\cite{Mindemark:2018gaa,Choo:2020iu,Popovic:2021jc} 

The crucial point to this conundrum lies in the fact that Bjerrum's convention is a thermodynamic criterion while the ionic conductivity is a dynamical property. Therefore, the lifetime of charge-neural ion-pairs needs to be considered explicitly when discussing the contribution of ion-pairing to the ionic conductivity, in addition to the distance criterion due to the thermodynamic stability. In other words, an ion-pair should be ``long-lived enough to be a recognizable kinetic entity''~\cite{1965_Robinson}.

Theoretically, the lifetime of ion-pairs $\tau_{+-}$ can be extracted from the normalized time correlation function of the cation-anion pairs in molecular dynamics (MD) simulations~\cite{2000_Luzar}:
\begin{equation}
\label{eq:res_corr}
        P(s) = \frac{\sum_i^N\sum_{j}^N\langle \theta(r_\textrm{c} - r_{ij}(0)) \cdot f(r_{ij};s)\rangle}
        {\sum_i^N\sum_{j}^N\langle \theta(r_\textrm{c} - r_{ij}(0)) \rangle}
\end{equation}
where $f(r_{ij};s)$ is a window function to detect whether a pair of cation-anion lies within the cutoff $r_c$ for a given period $s$. 

The first approach is to use the product of the Heaviside functions $\theta(x)$ defined
by a time series of pairwise distances $r_{ij}$ between a pair of cation-anion, as follows~\cite{Rapaport:1983ib}
\begin{equation}
\label{eq:f_rapaport}
f_\textrm{PT}(r_{ij};s) = \prod_{t<s} \theta(r_\textrm{c}-r_{ij}(t))
\end{equation}

However, the persistence time (PT) from this procedure clearly neglects 
recrossing events, e.g. reactions passing over the transition state but returning to the reactant afterwards, which has been discussed extensively for hydrogen bond
dynamics~\cite{1996_LuzarChandler,2000_Luzar}. Here, we used the stable
states picture (SSP) of chemical reactions proposed by Laage and Hynes, which remedies this problem.~\cite{2008_LaageHynes}. Then $f(r_{ij}; s)$ in SSP is given as: 
\begin{equation}
\label{eq:f_SSP}
    f_\mathrm{SSP}(r_{ij};s) = \prod_{t<s} \theta(r_\textrm{c,prod}-r_{ij}(t))
\end{equation}
where $r_\textrm{c,prod}$ is the product SSP boundary, which corresponds to the cation-anion distance at the half height of the second peak in the radial distribution function (RDF). Then, $r_\textrm{c}$ in Eq.~\ref{eq:res_corr} should be replaced by the reactant SSP boundary $r_\textrm{c,reac}$, which is at the first maximum of the cation-anion RDF.

To investigate the relation between  $\sigma^\textrm{d}_{+-}$ and $\tau_{+-}$ in polymer electrolyte systems, we constructed simulation boxes consisting of 200 PEO polymer chains each with 25 EO repeating units and 400 LiTFSI ion pairs ([Li$^+$]/[EO] concentration = 0.08). As indicated by Bjerrum's criterion, the solvent polarity strongly modulates the ion-pairing. This motivated us to apply the charge scaling method~\cite{Kirby:2019dm} to PEO molecules to change the degree of ion-pairing. General AMBER force field (GAFF)\cite{Wang2004} parameters were used for describing bonding and non-bonding interactions in PEO and LiTFSI and all molecular dynamics (MD) simulations were  performed using GROMACS 2018.1.\cite{Abraham2015}. All systems were properly equilibrated to make sure that the simulation length is larger than the Rouse time of the polymer. Details for the system setup and MD simulations can be found in the Supporting Information Section A.1.

Before discussing our main result, it is necessary to check how structural and transport properties change when we tweak the handle of the solvent polarity. Here, the solvent polarity is described by the dielectric constant of the system $\varepsilon_P$, which was computed for each polymer electrolyte system (See Section A.2 in the Supporting Information for details). 

\begin{figure}[h]
  \centering
  \includegraphics[width=1\linewidth]{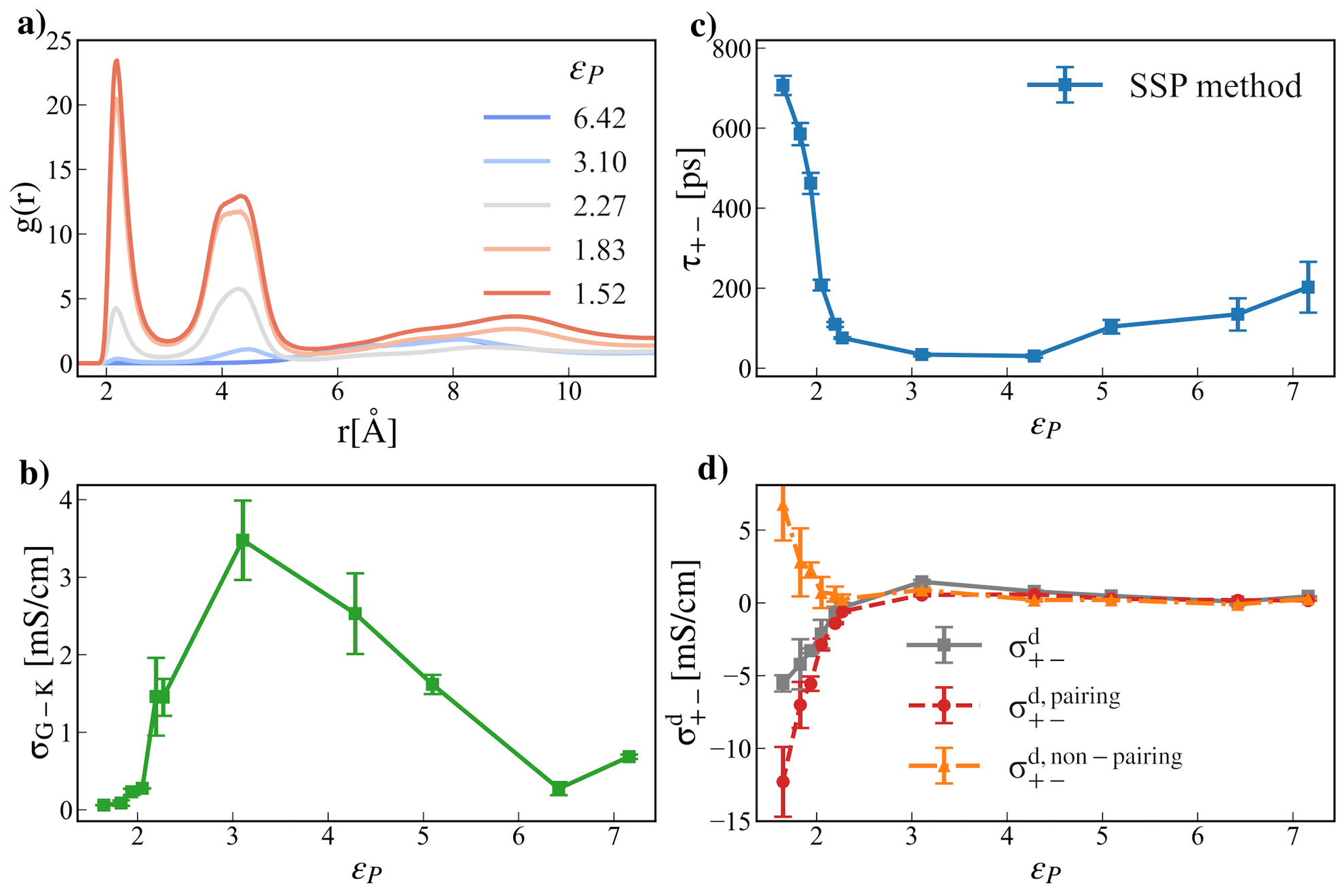}
  \captionsetup{font={stretch=1.5}}
  \caption{a) The Li-N(TFSI) radial distribution function at different solvent polarity strengths (as quantified by the dielectric constant $\varepsilon_P$); b) The total conductivity $\sigma_\textrm{G-K}$ computed from the Green-Kubo relation as a function of $\varepsilon_P$; c) The lifetime of ion-pairs $\tau_{+-}$ computed from the SSP method as a function of $\varepsilon_P$ where $r_\textrm{c,reac}= 2.1~$\AA~and $r_\textrm{c,prod}= 3.8 - 5.5~$\AA; d) The cation-anion distinct conductivity $\sigma^\textrm{d}_{+-}$ (and its decomposition into pairing and non-pairing contributions) as function of $\varepsilon_P$.}
  \label{fig:properties_vs_epsilonp}
\end{figure}

\begin{figure}[h]
  \centering
  \includegraphics[width=1.0\linewidth]{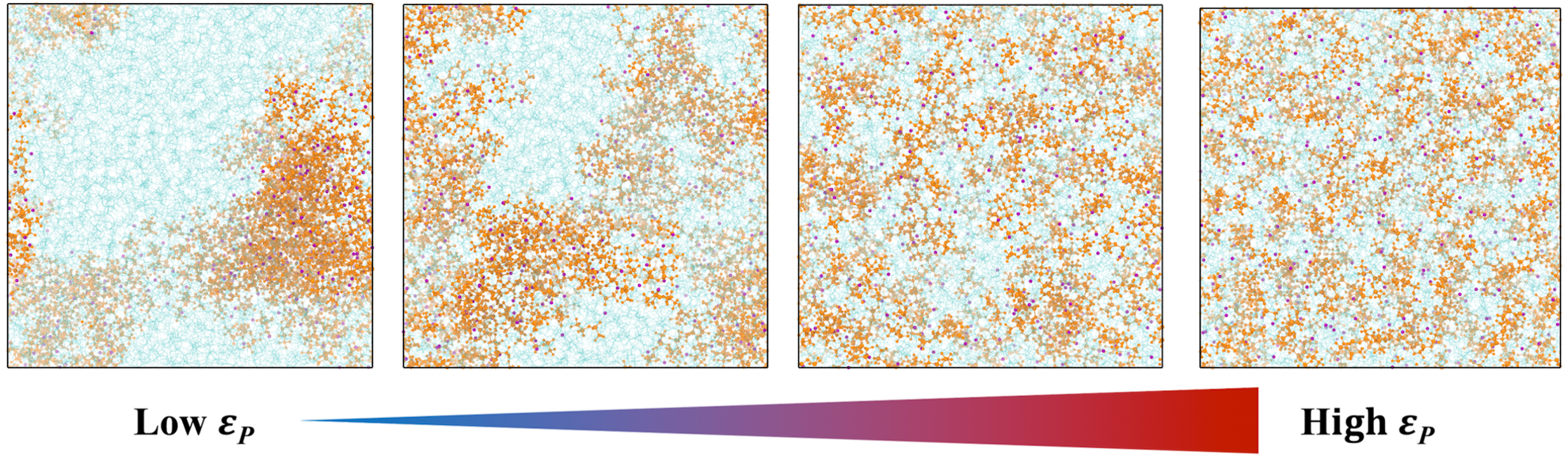}
  \captionsetup{font={stretch=1.5}}
  \caption{The modulation of ion-pairing in PEO-LiTFS systems by solvent polarity $\varepsilon_\textrm{P}$. Sky blue - PEO chains, Purple - Li-ions and Orange - TFSI ions.}
  \label{fig:ionpair_MD}
\end{figure}

The RDFs of Li-N(TFSI) are plotted in Fig.~\ref{fig:properties_vs_epsilonp}a, where peaks in the Li-N(TFSI) RDF increase significantly when $\varepsilon_P$ becomes smaller. This is a sign of formation of ion-pairs, which is also evinced in Fig.~\ref{fig:ionpair_MD}. Accordingly, there is an optimal value in the total Green-Kubo conductivity $\sigma_\textrm{G-K}$ when modulating the solvent polarity as seen in Fig.~\ref{fig:properties_vs_epsilonp}b. Both of these results support our previous observations of the effect of solvent polarity on the Li$^+$ transportation in PEO-LiTFSI systems~\cite{Gudla2020} and agree with other recent studies of polymer electrolyte systems~\cite{Wheatle:2018ed,Shen:2020jq}. 

Fig.~\ref{fig:properties_vs_epsilonp}c and Fig.~\ref{fig:properties_vs_epsilonp}d, on the other hand, demonstrate novel phenomena. The lifetime of ion-pairs increases when $\varepsilon_P$ is either high or low, and reaches a minimum at the intermediate value of $\varepsilon_P$ (See Section A.3 in the Supporting Information for further details of calculations of the lifetime of ion-pairs and comparison of outcomes from Eq.~\ref{eq:f_rapaport} and Eq.~\ref{eq:f_SSP}). Inspecting Fig.~\ref{fig:properties_vs_epsilonp}b and Fig.~\ref{fig:properties_vs_epsilonp}c, one may attempt to relate the opposite trend shown in the total ionic conductivity $\sigma_\textrm{G-K}$ to that of $\tau_{+-}$. However, the lifetime increases much more rapidly at lower dielectric constant regime ($\varepsilon_P < 2.3$) compared to that at higher dielectric constant regime ($\varepsilon_P > 3$). This suggests there are different types of ion-pairs in polymer electrolyte systems under investigation here. Looking at the cation-anion distinct conductivity $\sigma^\textrm{d}_{+-}$, one can clearly see that it goes from positive to negative when $\varepsilon_P$ becomes smaller (note that $\sigma^\textrm{d}_{+-}>0$ corresponds to anti-correlated cation-anion movements for the sign convention used in this work.). In particular, the rapid decrement in $\sigma^\textrm{d}_{+-}$ at lower $\varepsilon_P$ seems in accord with the rapid increment in $\tau_{+-}$. These observations also point to the direction that these two key properties of ion-pairs, namely $\sigma^\textrm{d}_{+-}$ and $\tau_{+-}$, must be closely related.

\begin{figure}[h]
  \centering
  \includegraphics[width=0.65\linewidth]{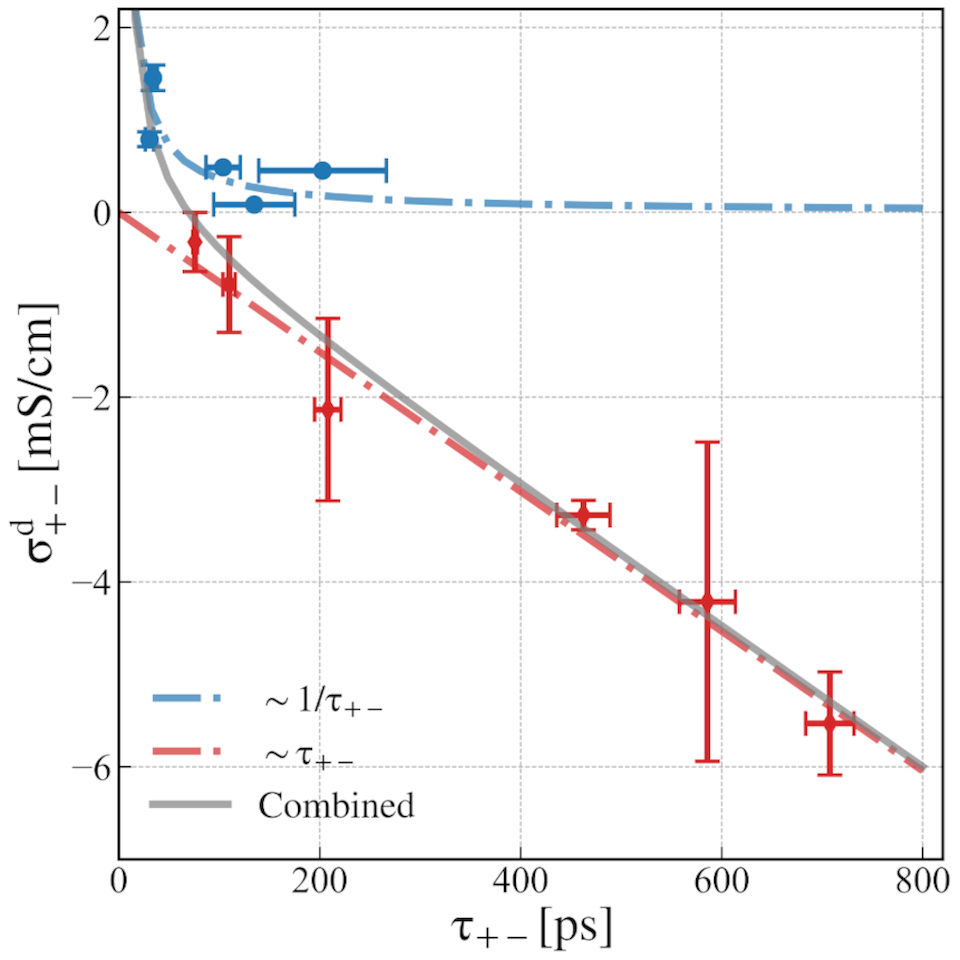}
  \captionsetup{font={stretch=1.5}}
  \caption{The scaling relation between the cation-anion distinct conductivity $\sigma^\textrm{d}_{+-}$ and the lifetime of ion-pairs $\tau_{+-}$ computed from the SSP method for PEO-LiTFSI polymer electrolyte systems with different solvent polarity strengths.}
  \label{fig:scaling_relation}
\end{figure}

This leads to our main result shown in Fig.~\ref{fig:scaling_relation}. What we find is that there exist two distinct regimes: $\sigma^\textrm{d}_{+-}$ scales with $1/\tau_{+-}$ (for higher values of $\varepsilon_P$) and $\sigma^\textrm{d}_{+-}$ scales with $\tau_{+-}$ (for  lower values of $\varepsilon_P$). Moreover, the transition between these two regimes shows a combined feature. Therefore, the general scaling relation we propose for polymer electrolyte systems is:
\begin{equation}
    \sigma^\textrm{d}_{+-} \sim \left(\frac{A}{\tau_{+-}} + B\cdot\tau_{+-}\right)
    \label{eq:scaling_relation}
\end{equation}
where both $A$ and $B$ are system-dependent coefficients. Therefore, what matters to discussions of the ion-pairing effect on transport properties in polymer electrolytes is not whether ion-pairs are present or not in the system but how long they live. By establishing the scaling relation for ion-pairs from MD simulations, one could predict the lifetime of ion-pairs using the measured value of $\sigma^\textrm{d}_{+-}$ in experiments~\cite{Pfeifer:2021ei}. 

Then, the immediate question appears why shorter-lived ion-pairs scale with $1/\tau_{+-}$ while longer-lived counterparts scale with $\tau_{+-}$? The first scaling relation seems rather general, as already observed in ionic liquids, organic electrolytes, polymer ionic liquids and salt-doped homopolymers~\cite{2015_ZhangMaginn,Mogurampelly:2017js,McDaniel:2018du, Shen:2020ke, Pfeifer:2021ei}. This is the reminiscent of the Walden rule or the Stokes-Einstein relation. The second scaling relation we find in this work is a consequence of that $\tau_{+-}$ computed from the SSP method equals to the inverse of the reactive flux rate constant $1/k_\textrm{RF}$~\cite{2008_LaageHynes} for the ion-pair dissociation and $1/k_\textrm{RF}$ is proportional to the concentration of ion-pairs from the law of mass action. Since $\sigma^\textrm{d}_{+-}$ is also proportional to the number of ion-pairs, this leads to the observation that $\sigma^\textrm{d}_{+-}$ scales linearly with $\tau_{+-}$. 

It is worth to mention that $\sigma^\textrm{d}_{+-}$ includes both contributions from the longer-lived ion-pairs and the remainder. This suggests that one could further separate these two contributions for longer-lived pairs:
\begin{equation}
  \label{eq:distinct_direct}
  \sigma^\textrm{d, pairing}_{+-} = \lim_{t \rightarrow \infty} \frac{1}{3tk_bT\Omega} \left[
  \sum_{i, +}  \sum_{j, -} \langle
  q_{i,+}q_{j,-}\Delta\mathbf{r}_{i,+}(t) \cdot \Delta\mathbf{r}_{j,-}(t)\cdot f_\textrm{SSP}(r_{ij}; s)
  \rangle \right] \\
\end{equation}
where $f_\textrm{SSP}$ is the same function given by Eq.~\ref{eq:f_SSP}. Then, the contribution from the remainder is simply $\sigma^\textrm{d, non-pairing}_{+-}=\sigma^\textrm{d}_{+-}-\sigma^\textrm{d, pairing}_{+-}$. Here, the parameter $s$ is chosen to be 2~ns, as decided by the convergence of the conductivity calculation (See Section A.4 in the Supporting Information).

The result of this decomposition is shown in Fig.~\ref{fig:properties_vs_epsilonp}d. $\sigma^\textrm{d, pairing}_{+-}$ remains zero until a lower value of $\varepsilon_P$. This agrees with the appearance of longer-lived ion-pairs as seen in Fig.~\ref{fig:properties_vs_epsilonp}c. More interestingly, in the presence of longer-lived ion-pairs,  $\sigma^\textrm{d, pairing}_{+-}$ is negative, but $\sigma^\textrm{d, non-pairing}_{+-}$ is positive instead. To understand why, we made a toy model of NaCl solution where all Na-Cl are paired up with holonomic constraints. Details for the system setup and MD simulations can be found in Section B of the Supporting Information. 

\begin{figure} [h]
  \centering
  \includegraphics[width=0.65\linewidth]{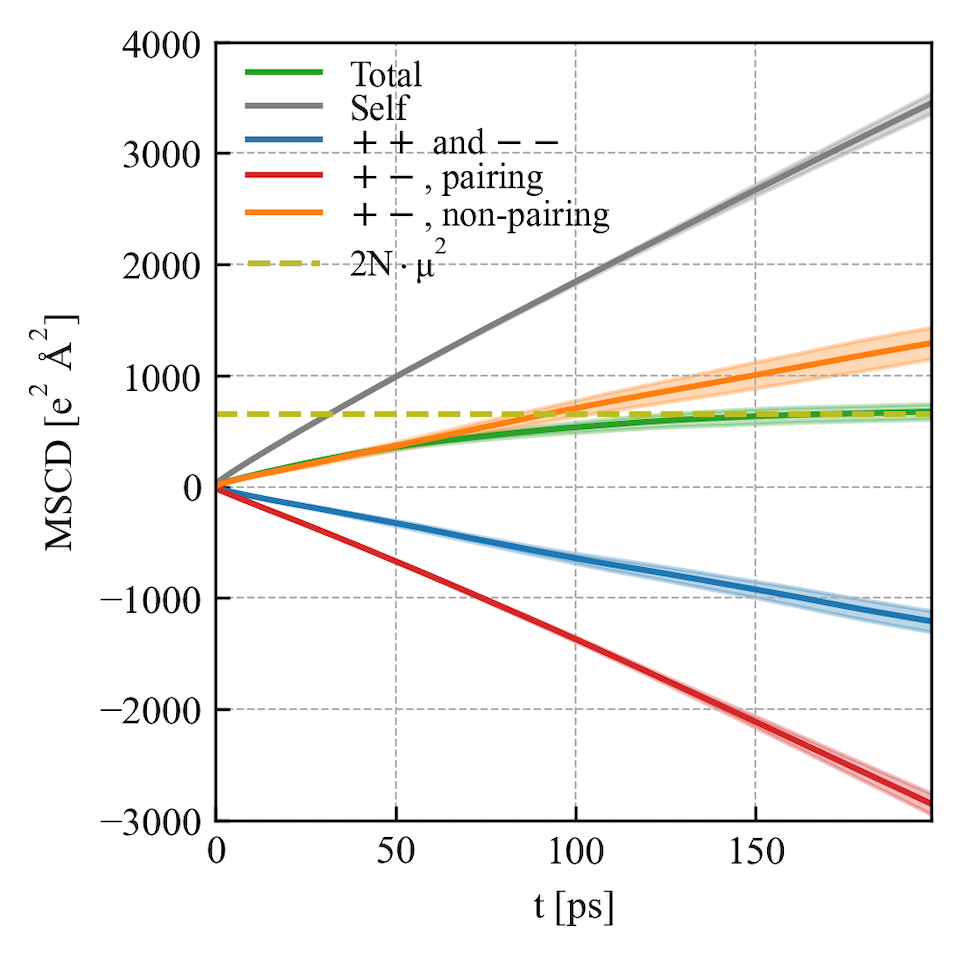}
  \captionsetup{font={stretch=1.5}}
  \caption{Mean square charge displacements (MSCD) of a 5 mol NaCl solution with permanent ion-pairs (in the same order as those appeared in the key box): for the total ionic conductivity $\sigma_\textrm{G-K}$, the self-conductivity $(\sigma_{+} + \sigma_{-})$, the sum of cation-cation and anion-anion distinct conductivities $(\sigma^\textrm{d}_{++} + \sigma^\textrm{d}_{--})$, the ion-pairing part of the cation-anion distinct conductivity $\sigma^\textrm{d, pairing}_{+-}$ and the remainder part $\sigma^\textrm{d, non-pairing}_{+-}$. $N$ is the number of ion-pairs in the system and $\mu$ is the dipole moment of each ion-pair. }
  \label{fig:nacl_MSCD}
\end{figure}

Mean square charge displacements (MSCD, i.e. quantities inside the square bracket in Eq.~\ref{eq:distinct conductivity} and Eq.~\ref{eq:distinct_direct}) of this toy model are shown in Fig.~\ref{fig:nacl_MSCD}, for the total ionic conductivity $\sigma_\textrm{G-K}$,  self-conductivities $(\sigma_{+} + \sigma_{-})$, the sum of cation-cation and anion-anion distinct conductivities $(\sigma^\textrm{d}_{++} + \sigma^\textrm{d}_{--})$, as well as $\sigma^\textrm{d, pairing}_{+-}$  and $\sigma^\textrm{d, non-pairing}_{+-}$. Since all Na-Cl ion-pairs are permanent by construction, the total ionic conductivity as the sum of all these individual contributions mentioned above must be zero (i.e. the slope of MSCD ``total'' is zero), as evinced in Fig.~\ref{fig:nacl_MSCD}. Moreover, self-conductivities $(\sigma_{+} + \sigma_{-})$ should be exactly the negative of the direct part of the cation-anion distinct conductivity $\sigma^\textrm{d, pairing}_{+-}$, as seen also in Fig.~\ref{fig:nacl_MSCD}. Based on these considerations, we know that the sum of $\sigma^\textrm{d}_{++}$, $\sigma^\textrm{d}_{--}$ and $\sigma^\textrm{d, non-pairing}_{+-}$ is zero as well. As shown in Fig.~\ref{fig:nacl_MSCD}, $\sigma^\textrm{d}_{++}$ and $\sigma^\textrm{d}_{--}$ are negative while $\sigma^\textrm{d, non-pairing}_{+-}$ is positive in the toy model with permanent ion-pairs. This provides a rationale to the opposite signs of $\sigma^\textrm{d, pairing}_{+-}$ and $\sigma^\textrm{d, non-pairing}_{+-}$ as seen in Fig.~\ref{fig:properties_vs_epsilonp}d of PEO-LiTFSI systems. Nevertheless, one should be aware that the situation with ion aggregates will be different, as $\sigma^\textrm{d}_{++}$ and $\sigma^\textrm{d}_{--}$ could be positive instead.

To sum up, following Bjerrum's criterion, we have constructed PEO-LiTFSI systems in our MD simulations with different degree of ion-pairing by modulating the solvent polarity. What we found is that there exist two distinct regimes where the cation-anion distinct conductivity $\sigma^\textrm{d}_{+-}$ scales with $1/\tau_{+-}$ and with $\tau_{+-}$ respectively. The linear scaling of $\sigma^\textrm{d}_{+-}$  with respect to the lifetime of the ion-pairs $\tau_{+-}$ is a signature of longer-lived ion-pairs which reduce the total ionic conductivity. By establishing this scaling relation, one could infer the lifetime of ion-pairs from the experimentally measured  cation-anion distinct conductivity. This further suggests that what matters to discussions of the ion-pairing effect on transport properties in polymer electrolytes is not the presence of ion-pairs but the corresponding lifetime. 

In the scaling relation we found in our MD simulations (Eq.~\ref{eq:scaling_relation}), the coefficient $A$ is positive which suggests anti-correlated movements of cation-anions and shorter-lived ion-pairs. This hints that ion aggregates analyzed in previous MD studies of the PEO-LiTFSI system~\cite{Molinari18} would not populate in this scenario, in line with conclusions drawn from other experimental works for polymer electrolytes.~\cite{REY19981505,Popovic:2015kn} Nevertheless, it is worth to note that early experiments on aqueous ionic solutions show that $\sigma^\textrm{d}_{+-}$ can flip the sign from negative (correlated) to positive (anti-correlated) when increasing the salt concentration~\cite{Woolf:1978db}, which is intriguing. This clearly indicates that the cation-anion distinct conductivity $\sigma^\textrm{d}_{+-}$ is a sensitive probe to the convoluted ion-ion correlations, which calls for further investigations from both experiments and simulations to understand its nature and its relationship with other static and dynamical properties in electrolyte systems.

\begin{suppinfo}
Descriptions of the setup and MD simulations of PEO-LiTFSI systems and NaCl solution with permanent ion-pairs.
\end{suppinfo}

\begin{acknowledgement}
  This work has been supported by the European Research Council (ERC), grant no.
  771777 “FUN POLYSTORE” and the Swedish Research Council (VR), grant no.  2019-05012. The authors thanks funding from the Swedish National
  Strategic e-Science program eSSENCE and STandUP for Energy. The simulations were performed on the
  resources provided by the Swedish National Infrastructure for Computing (SNIC)
  at NSC, PDC and UPPMAX.
\end{acknowledgement}

\providecommand{\latin}[1]{#1}
\makeatletter
\providecommand{\doi}
  {\begingroup\let\do\@makeother\dospecials
  \catcode`\{=1 \catcode`\}=2 \doi@aux}
\providecommand{\doi@aux}[1]{\endgroup\texttt{#1}}
\makeatother
\providecommand*\mcitethebibliography{\thebibliography}
\csname @ifundefined\endcsname{endmcitethebibliography}
  {\let\endmcitethebibliography\endthebibliography}{}


\end{document}


\maketitle

\makeatletter 
\renewcommand{\thefigure}{S\@arabic\c@figure} \renewcommand{\thetable}{S\@arabic\c@table}
\renewcommand{\theequation}{S\@arabic\c@equation} 
\renewcommand{\thepage}{S\arabic{page}}

\makeatother

\appendix{} 
\setcounter{figure}{0} 
\setcounter{table}{0}

\centerline{\bf\Large{Supplementary Information}}

\newpage 

\section{Description of the setup and MD simulations of PEO-LiTFSI systems}
\subsection{Force field and simulation details}
General AMBER force field (GAFF)\cite{Wang2004} parameters were used for describing bonding and non-bonding interactions in PEO and LiTFSI and all simulations were  performed using GROMACS 2018.1.\cite{Abraham2015} 
\begin{table}[h!]
\centering
 \begin{tabular}{||c c c c||} 
 \hline
 \multicolumn{2}{||c}{PEO} & \multicolumn{2}{c||}{LiTFSI}\\
 \hline
 Atom type & Charge (e) & Atome type & Charge (e) \\ [0.5ex] 
 \hline\hline
 C & 0.13 & C & 0.41 \\ 
 O$_\textrm{chain}$ & -0.43 & S & 1.11, 1.26 \\
 H$_\textrm{chain}$ & 0.04 & F & -0.20 \\
 O$_\textrm{end}$ & -0.61 & N & -0.79 \\
 H$_\textrm{end}$ & 0.41 & O & 0.46,0.49 \\
  &  & Li & 0.75 \\ [1ex] 
 \hline
 \end{tabular}
 \caption{Intial partial charges of PEO from GAFF and scaled charges of LiTFSI from GAFF.}
\label{tab:gaff_charges}
\end{table}

MD simulation boxes comprising neat and LiTFSI-doped PEO systems were constructed, comprising 200 hydroxyl-terminated chains, each with 25 monomer units (1.11 kg/mol) and 400 Li and 400 TFSI ions, corresponding to a [Li+]/[EO] concentration ratio of 0.08.
All systems were equilibrated using a Bussi-Donadio-Parrinello\cite{bussi_2007} thermostat and a Parrinello-Rahman barostat\cite{Parrinello1981} at 400 K and 1 bar for 5 and 10 ns, respectively with a time step of 1 fs and snapshots saved every 5 ps. Then, NPT (constant number of particles, constant pressure, and constant temperature) production runs were carried out for additional 400 ns at a desired temperatures. The specific temperatures were chosen to separate the effect of glass transition temperature and solvent polarity on the polymer and ion dynamics.\cite{Gudla2020} The dielectric constant of the system was estimated from the fluctuations in the total dipole moment of the PEO was used as a measure of the solvent polarity.  The solvent polarity of the system can be modulated by scaling the partial atomic charges on polymer. The scaling factor and dielectric constant of different systems used in this work is reported in Table~\ref{tab:sim_details}.

\begin{table}[h!]
\centering
 \begin{tabular}{||c c c c||} 
 \hline
 Scaling factor & T (K) & $\varepsilon_P$ & box size (nm$^3$) \\ [0.5ex] 
 \hline\hline
 1.55 & 450 & 7.16 (0.22) & 438.33 \\ 
 1.50 & 450 & 6.42 (0.11) & 440.58 \\
 1.34 & 440 & 5.09 (0.03) & 442.77 \\
 1.20 & 440 & 4.29 (0.02) & 447.30 \\
 1.00 & 430 & 3.10 (0.02) & 450.29 \\
 0.75 & 420 & 2.27 (0.05) & 453.97 \\
 0.70 & 420 & 2.19 (0.00) & 455.04 \\
 0.65 & 420 & 2.05 (0.02) & 456.78 \\
 0.60 & 410 & 1.94 (0.00) & 454.98 \\
 0.56 & 410 & 1.83 (0.03) & 456.76 \\
 0.50 & 410 & 1.65 (0.00) & 458.74 \\ [1ex] 
 \hline
 \end{tabular}
 \caption{Details of the different simulated systems.}
\label{tab:sim_details}
\end{table}

\subsection{Calculation of the dielectric constant of polymer electrolyte systems}

The dielectric constant of the polymer electrolyte $\varepsilon_P$ reflects the strength of solvent polarity. The dielectric constant of PEO-LiTFSI system with periodic boundary conditions and Ewald summation can be computed from the fluctuations in the total dipole moment $\mathbf{M}$ of the PEO via the equation for a non-polarizable model\cite{Neumann1983}:

\begin{equation}
    \varepsilon_p=1+\ \frac{4\pi}{3\varepsilon_0\Omega k_BT}(\langle \mathbf{M}^2\rangle - \langle \mathbf{M}\rangle^2)
\end{equation}

where $\varepsilon_0$ is vacuum permittivity, $k_B$ is Boltzmann’s constant, $T$ is temperature, $\Omega$ is the average volume of the simulation box and $\langle\ \rangle$ indicates ensemble averages. 

\subsection{Calculation of lifetime of ion-pairs and comparison of methods }
The lifetime of ion-pairs can be extracted from the normalized time correlation function representing the life
expectancies of the pairs, i.e. the first passage time probability function\cite{2000_Luzar}:

\begin{equation}
\label{eq:res_corr}
    P(s) = \langle \theta(r_\textrm{c} - r_{ij}(0)) \cdot f(r_{ij};s)\rangle/\langle \theta(r_\textrm{c} - r_{ij}(0)) \rangle
\end{equation}
where $f(r_{ij};s)$ can be calculated from Eq. 3 for persistence lifetime ($\tau_{+-}^\textrm{PT}$) or from Eq. 4 for stable state picture lifetime ($\tau_{+-}^\textrm{SSP}$) (See the Main Text for Eq.3 and Eq.4). The lifetime can then be obtained by fitting $P(s)$ to a biexponentially decaying
function which converges to a constant.
\begin{align}
      P(s) &= (1-A_1-A2)+A_1 \exp{(-\frac{s}{\mathrm{\tau_1}})} + A_2 \exp{(-\frac{s}{\tau_2})}  \\
      \tau_{+-} &= \frac{A_1\tau_1 + A_2\tau_2}{A_1+A_2}
\end{align}

 The  normalized  time  correlation function ($P(s)$) of Li and N(TFSI) were calculated for a correlation time of 5~ns for every 20~ns segment of the trajectory (See Fig. S1(a)). The ion-pair lifetime values from both the methods $\tau_{+-}^\textrm{PT}$ and $\tau_{+-}^\textrm{SSP}$ were plotted in Fig. S1(b), we can observe that the persistence lifetimes were much lower than the SSP lifetime because of missing out the recrossing events. 
 
 \begin{figure}[h]
  \centering
  \includegraphics[width=1\linewidth]{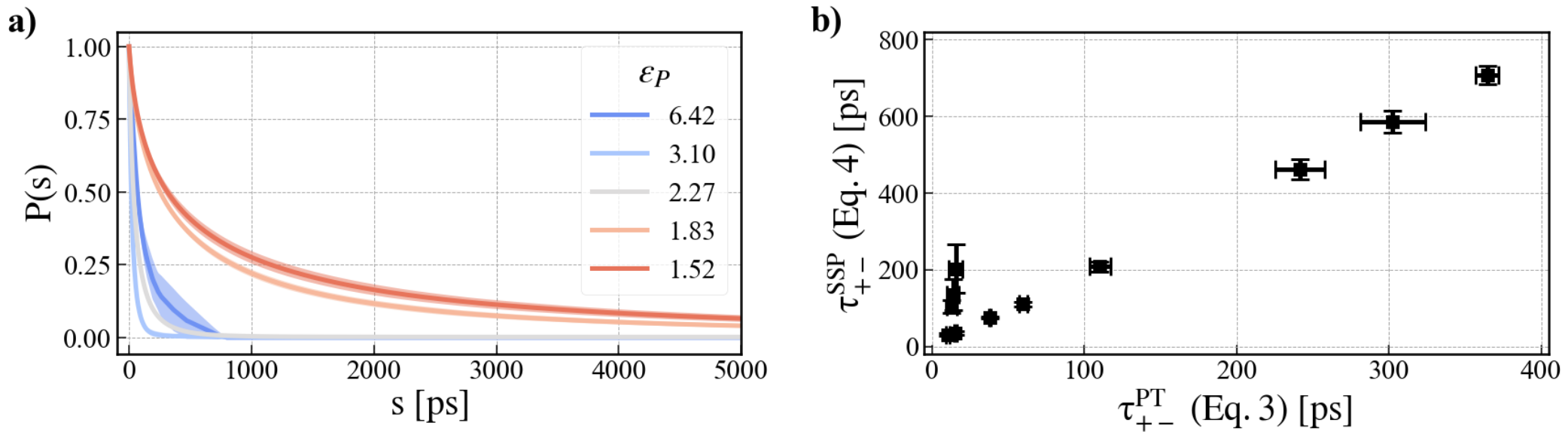}
  \caption{a) The normalized time correlation function ($P(s)$) of Li and N(TFSI) for sytems with differnt scaling factors or solvent polarity ($\varepsilon_P$). b) The pesistance lifetime ($\tau_{+-}^\textrm{PT}$) vs stable state picture lifetime ($\tau_{+-}^\textrm{SSP}$) at different solvent polarities. }
  \label{fig:ps_ptvsssp}
\end{figure}

\subsection{Convergence of the conductivity calculations}

The Green-Kubo (G-K) formula that was used to calculate the total ionic conductivity is:
\begin{eqnarray}
  \label{eq:G-K conductivity}
  \sigma_{\textrm{G-K}} &=& \lim_{t \rightarrow \infty} \frac{1}{6tk_bT\Omega} \left[
  \sum_i^N  \sum_{j}^N \langle
  q_iq_j\Delta\mathbf{r}_{i}(t) \cdot \Delta\mathbf{r}_{j}(t)
  \rangle \right]
\end{eqnarray}
where $\Omega$ is the volume of the system and $\Delta\mathbf{r}(t)$ is the displacement vector of each ion at time $t$. The cumulative averages of instantaneous ionic conductivity values of $\sigma_{\textrm{G-K}}$ and $\sigma_{+-}^d$ were used to calculate the reported values and were plotted for $\varepsilon_P=\ 2.19$ in Fig. S2(a). For all the systems, these values converges after 100~ns. Therefore, the final values were calculated by averaging over 120~ns to 140~ns. The convergence of $\sigma_{+-}^{\textrm{d,pairing}}$ with $r_c$ of 3.1~Å was also checked and plotted in Fig. S2(b) for PEO-LiTFSI system with different solvent polarities. The final values of $\sigma_{+-}^{\textrm{d,pairing}}$ were calculated by averaging over 1.4~ns to 1.8~ns.
 \begin{figure}[h]
  \centering
  \includegraphics[width=1\linewidth]{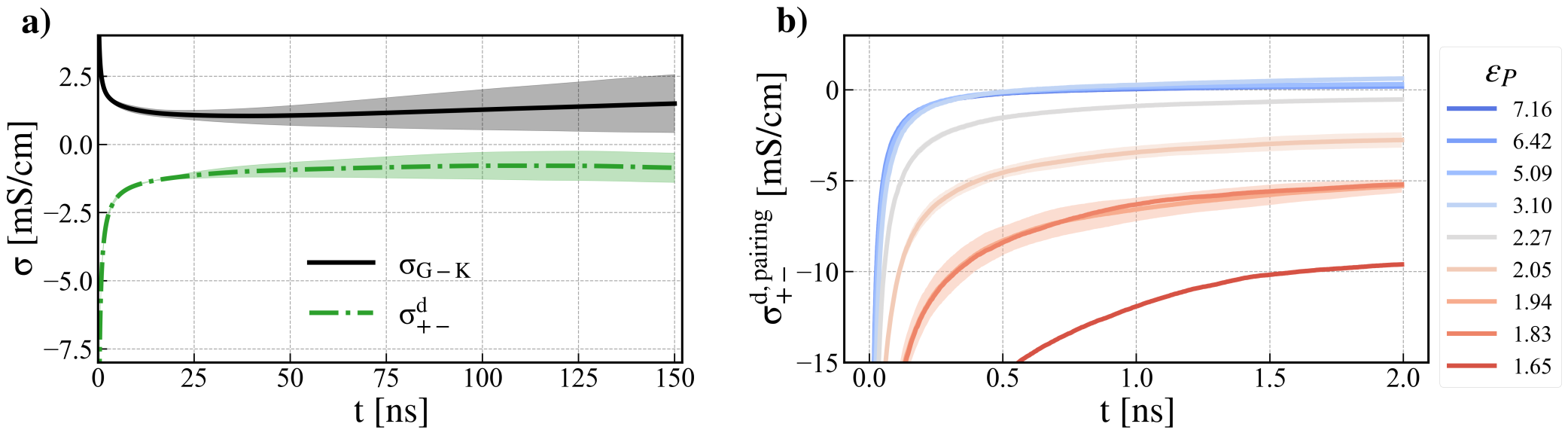}
  \caption{a) The instantaneous ionic conductivity values of $\sigma_{\textrm{G-K}}$ and $\sigma_{+-}^d$ for the PEO-LiTFSI system with $\varepsilon_P=\ 2.19$. b) The instantaneous $\sigma_{+-}^{\textrm{d,pairing}}$ for the PEO-LiTFSI sytem at different solvent polarities where ion-pairs are defined with a cutt-off distance ($r_c$) of 3.1~Å and time period ($s$) of 2~ns. }
  \label{fig:ps_ptvsssp}
\end{figure}

\section{Description of the setup and MD simulations of NaCl solution with permanent ion-pairs}

Systems of sodium chloride electrolyte solution where the water molecules were
described with the simple point charge/extended (SPC/E) model\cite{berendsen_1987} and Na/Cl ions are described with parameters proposed by Joung and Cheatham\cite{joung_2008}. To introduce permanent ion-pairs to the system, holonomic constraints through the SHAKE algorithm~\cite{Ryckaert:gp}  were added between selected pair of ions, the bond length was set to be 2.85~\AA. according to the position of the first peak in the Na-Cl RDF in the unconstrained system.
The size, density and number of molecules used in the simulation are listed in Table~\ref{tab:stoichiometry_nacl}.

\begin{table}[htb]
    \centering
    \begin{tabular}{||ccccc||}
    \hline
         $m$ [mol kg$^{-1}$] &  $\rho$ [g cm$^{-3}$] 
         &  $N_\mathrm{NaCl}$  &  $N_{\mathrm{H_2O}}$  & $L$ [\AA] \\
    \hline
    \hline
        5.14 & 1.17 & 40 &  432 & 24.29\\
    \hline
    \end{tabular}
    \caption{Size and composition of the MD simulations of NaCl solution in this work: molality $m$, density $\rho$, number of NaCl and H$_2$O molecules  $N_\mathrm{NaCl}$ and $N_\mathrm{H_2O}$ and length of the cubic simulation box $L$.}
    \label{tab:stoichiometry_nacl}
\end{table}

The molecular dynamics simulations were performed with the LAMMPS code. The
long-range electrostatics were computed using the particle-particle particle-mesh
(PPPM) solver\cite{hockney_1988}. Short-range cutoffs for the van der Waals and
Coulomb interactions in direct space were 9.8~\AA. For each system, NVT (constant
number of particles, constant volume, and constant temperature) simulations ran
for 20~ns with a timestep of 2~fs and trajectories were collected every 0.5~ps.
The Bussi-Donadio-Parrinello thermostat\cite{bussi_2007} was used to maintain
the given temperature of 293~K.
